\documentclass[english,aps,preprint]{revtex4}
\usepackage[T1]{fontenc}
\usepackage[latin1]{inputenc}
\usepackage{amsmath}
\usepackage{graphicx}
\usepackage{amssymb}

\makeatletter
\usepackage{babel}
\makeatother
\begin{document}

\preprint{Submitted to American Journal of Physics 2/10/07, Resubmitted with
revisions 6/28/07}

\title{Two-Particle Models, Diagrams and Graphs as Aids to Understanding
Internal Energy}

\author{Matthew L. Greenwolfe}

\email{matt_greenwolfe@caryacademy.org}

\affiliation{Cary Academy, 1500 N. Harrison, Ave., Cary, NC 27513}

\begin{abstract}
Single particle models are sufficient for many topics in introductory
physics courses, but become misleading when applied to situations
involving changes in internal energy. This paper merges two parallel
attempts to reform the teaching of energy in the introductory physics
curriculum, resulting in an approach to internal energy that is conceptually
correct, mathematically rigorous, and yet simple enough for the introductory
student to understand. It uses diagrams and graphs to provide concrete
and visual means to track energy storage and transfer, and takes the
smallest possible step in mathematical complexity - from one-particle
to two-particle models - necessary to achieve consistency with a sound
conceptual approach. The mathematical development is guided by the
idea that the term {}``work'' should receive a clear conceptual
definition as energy that is transferred through the boundary of a
system; mathematical expressions are identified as work - of any sort
- only when they correspond to this definition. 
\end{abstract}

\keywords{energy, work, diagram, graph, model}

\pacs{01.40.gb Teaching methods and strategies 45.20.dg Mechanical energy,
work, and power 45.20.dh Energy conservation}

\maketitle

\section{introduction}

This paper merges two parallel attempts to reform the teaching of
energy in the introductory physics curriculum, one more mathematical,
and the other more conceptual and visual. Many authors have pointed
out that single particle mathematical models become misleading when
applied to situations involving changes in internal energy.\cite{alonso-finn,arons,sherwood-pseudowork}
Such situations are a valuable lever to induce reform of the physics
curriculum, because they dramatically make a crucial point: that a
student's ability to obtain a correct mathematical answer to a problem
has little correlation with conceptual understanding. In a recent
TPT article,\cite{Mungan-review} Carl Mungan reviewed proposals for
dealing with internal energy. In a much earlier review, Malinckrodt
and Leff\cite{mall-leff-all-about-work} identified a total of seven
different types of work referenced in the literature, which are related
to six state-dependent energy functions, producing a multiplicity
of work-energy relationships which they selectively apply to solve
problems. As Mallinckrodt and Leff say, {}``The abundance of papers
devoted to energy transformations, and in particular, work in macroscopic
systems is indicative of the discomfort many physics teachers experience
in the area.''

These attempts to reform the energy curriculum have in common a desire
to formulate the concepts, terminology, and mathematics of internal
energy to dovetail as seamlessly as possible with the typical treatment
of the subject in textbooks and introductory courses. This creates
an important bridge between elementary mechanics on the one hand,
and thermodynamics and solid-state physics on the other. Since the
energy curriculum in introductory physics typically starts with the
single-particle work-energy theorem, these authors spend considerable
effort to delineate the scope of this theorem so that teachers and
students can identify the class of problems to which it applies, and
the class of problems for which it fails. Since the typical curriculum
defines work using a dot product of force and displacement of the
center of mass, these authors recommend terms such as {}``pseudo-work,''
{}``center-of-mass-work,'' and {}``particle-work'' to refer to
this product when it does not equal an amount of energy transferred
through the boundary of a system - which is the only thermodynamically
valid definition of work. They apply both a multi-particle first law
of thermodynamics and a single-particle work-energy theorem to the
same physical system, and solve the equations simultaneously.\cite{work-energy-only-dyn}

Others argue that it's better to teach introductory students to qualitatively
track the storage and flow of energy, leaving nuanced variations in
terminology and complex mathematics to more advanced courses.\cite{swartz}
Indeed, another emerging reform of the introductory energy curriculum
uses qualitative and semi-quantitative diagrams and graphs as developed
by vanHeuvelen and Zou\cite{vanHeuvelen-zou}, Falk and Hermann\cite{falk-herrmann},
and adapted by the Modeling Workshop Project\cite{modeling-teacher-notes}.
In this approach, diagrams and graphs provide a concrete and visual
means to track energy storage and transfer, supplementing verbal descriptions
and equations. Using this approach, the course sequence is substantially
changed. The first law of thermodynamics is introduced from the very
beginning by defining energy as {}``the ability to cause change.''
This definition is more general than {}``the ability to do work,''
because it does not restrict energy to mechanics, but includes energy
transfer by conducting and radiating as well as working. Students'
first exercises in the subject require them to demonstrate their understanding
of energy conservation by using diagrams and graphs to show where
and how energy is stored at every instant of time. These qualitative
and semi-quantitative exercises include energy storage in biological,
chemical, and thermal, as well as mechanical systems.\cite{unitary-energy}
The diagrams and graphs also require students to explicitly identify
the objects inside their system and to show the transfer of energy
into or out of that system, in correspondence with the thermodynamic
definition of work. Since the curriculum includes internal energy
stored in multi-particle systems from the beginning, invocation of
the single-particle work energy theorem is unnecessary and is therefore
frequently deleted. Instead, students solve quantitative problems
by matching formulas for energy storage or transfer with different
parts of the diagrams. The structural relationship among these parts
serves as a guide to develop a conservation of energy equation (the
first law of thermodynamics) specific to the situation. 

The mathematical formalism of Mungan, Sherwood, Arons et. al. dovetails
nicely with the typical textbook presentation of energy, but fits
awkwardly with the qualitative-visual approach of Swartz, vanHeuvelen,
Falk, Swackhamer and others. In the qualitative-graphical approach,
the work-energy equation has not been introduced in the first place,
so there is little need to spend time explaining its limits, nor is
it convenient to use it for mathematical solutions in combination
with the first law of thermodynamics. Since work is defined as thermodynamic
work and given explicit visual representation, such terms as {}``pseudo-work,''
{}``center-of-mass-work,'' and {}``internal-work'' become misleading.
I find great value in both the qualitative-visual approach and the
rigor of the mathematical formalism. In this paper, I show how these
two approaches can exist together more harmoniously.

Since the curriculum is aimed at introductory students, this paper
takes the smallest possible step in mathematical complexity - from
one-particle to two-particle models - necessary to achieve consistency
with a sound conceptual approach to energy. The mathematical development
is guided by the idea that the term {}``work'' should receive a
clear conceptual definition as energy that is transferred through
the boundary of a system; mathematical expressions are identified
as work - of any sort - only when they correspond to this definition.
Even greater simplification is achieved by extracting some physically
sensible guidelines from the mathematical theory, enabling introductory
students to bypass difficult mathematics while making sense of the
most troublesome examples. The result is an approach to these issues
that is conceptually correct, mathematically rigorous, and yet simple
enough for the introductory student to understand. It is mathematically
equivalent to previous formulations, but obtains the advantage that
the mathematics flows directly from the conceptual analysis in a straightforward
manner.

\section{Introducing internal energy concepts using diagrams and graphs}

In this section, I review a collection of problems for which changes
in internal energy play a prominent role. These problems have been
discussed extensively by previous authors \cite{arons,sherwood-pseudowork,swackhamer-work-work}
who show that the conflict between a single particle mathematical
model and the energy concepts creates a barrier to student understanding
of energy. My main purpose is to show how students can use pie charts,
bar graphs and energy flow diagrams to illustrate the storage and
transfer of energy. 

\begin{figure}
\includegraphics{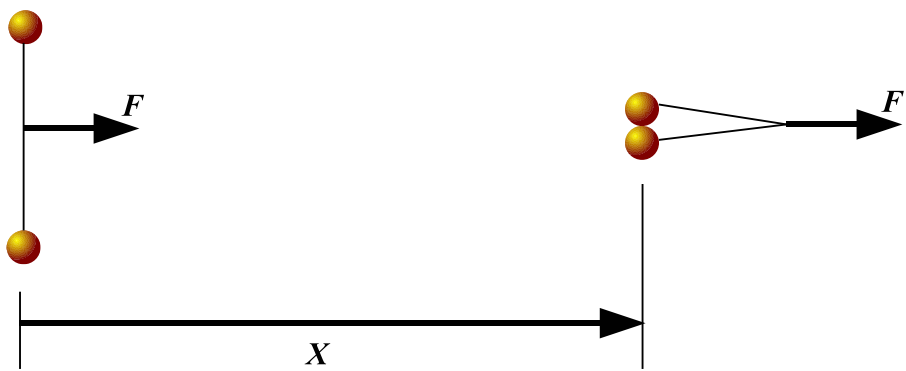}

\caption{\label{fig:2pucks-string}two frictionless pucks of mass m, connected
by a string, are accelerated by the force F.}
\end{figure}

Bruce Sherwood attributes the following problem to Michael Wiessman.\cite{sherwood-pseudowork}
It is also extensively discussed by Arnold Arons.\cite{arons-book}
I adopt Aron's description. {}``Consider the situation {[}in figure
\ref{fig:2pucks-string},] in which two frictionless pucks of mass
$\mathit{m}$, connected by a string, are accelerated by the force
$\mathit{F}$. (Note that this is a deformable system, that the force
is displaced farther than the center of mass point, and that the pucks
are assumed to undergo an inelastic collision on contact.)''

\begin{figure}
\includegraphics{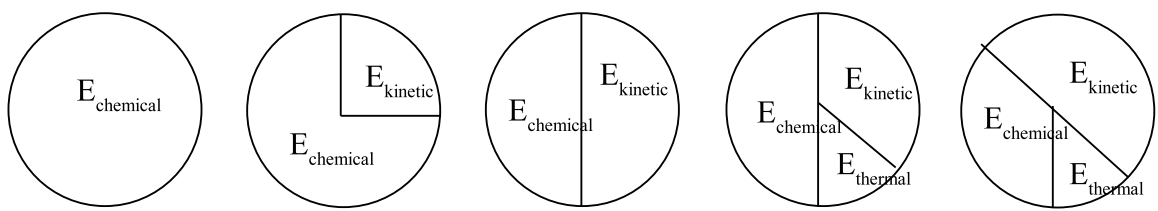}

\caption{\label{fig:2pucks-pie-charts}Pie charts illustrating that energy
starts as chemical energy in the person's muscles and is gradually
transferred to the pucks and stored as kinetic energy and thermal
energy.}
\end{figure}

By the time they arrive in an introductory physics course, most students
can glibly recite a version of energy conservation such as, {}``Energy
can neither be created nor destroyed.'' Energy pie charts (Figure
\ref{fig:2pucks-pie-charts}) require students to express their understanding
of energy conservation in a stronger form. The pie charts track an
amount of energy over time as it is transferred from one location
to another, requiring students to identify where and how the energy
is stored at every instant of time. If there is any instant when they
cannot do so, this points out a deficiency in their energy model in
violation of conservation of energy. At first, students have difficulty
drawing a correct set of energy pie charts to describe even a simple
situation because they typically harbor persistent misconceptions
about the meaning of energy and its conservation, even when they can
recite a version of energy conservation and correctly solve the problem
mathematically.\cite{arons,Hilborn,swackhamer-work-work} In my classroom,
I often probe students' understanding by asking them to draw pie charts
before, after, or between the ones they have already drawn, inevitably
revealing beliefs that kinetic energy is {}``being used'' and is
not stored anywhere, that transfer of energy is not a gradual process
but happens all at once {}``from potential to kinetic,'' and a belief
in the equivalence of rest and potential energy storage, so that falling
objects instantly recover their gravitational energy upon impact with
the ground. If you don't believe your students harbor misconceptions,
try asking them to draw a set of pie charts. Figure 2 shows that the
energy that eventually ends up in the pucks starts out stored in the
person's muscles as chemical potential energy and is gradually transferred
to the pucks, where it is stored partially as kinetic energy and partially
as thermal energy because of the inelastic collision.\cite{w-e-2puck}

\begin{figure}
\includegraphics{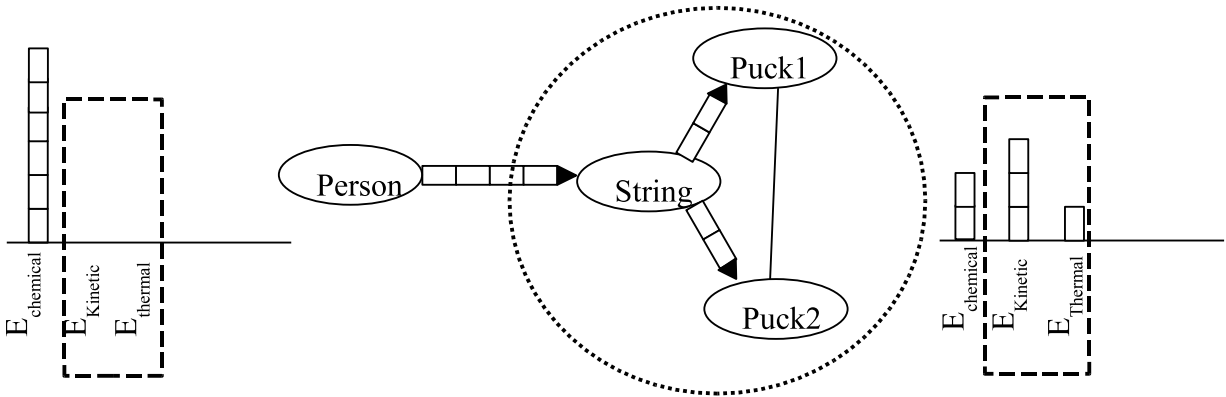}

\caption{\label{fig:bar-flow-2puck}Energy bar graphs and energy flow diagram
showing similar information to the pie charts, but emphasizing the
choice of system and its role in defining work.}
\end{figure}

Energy bar graphs and energy flow diagrams (figure \ref{fig:bar-flow-2puck})
together show much the same information as the energy pie charts,
but with a different emphasis. In both representations, the dotted
lines demarcate the boundaries of the system. Only objects and energy
inside the boundary are actually in the system. In the energy flow
diagram, a line indicates an interaction between the objects. Adding
an arrow to the line indicates flow of energy. Clearly defining a
system is a crucial step in solving any physics problem, and energy
is no exception. These diagrams make the separation between system
and environment explicit. The bar graphs focus student attention on
the initial and final energy state of the system, and require students
to consciously consider conservation of energy by making their energy
bars stack up to the same height. By contrast, each pie chart automatically
represents the same amount of energy, forcing student ideas to be
consistent with energy conservation even if they haven't given the
matter explicit thought. The energy flow diagram focuses students
on the objects where the energy is stored and the direction of energy
transfer. The pie charts, by contrast, focus more on the mechanism
of energy storage - chemical, kinetic, etc. - and imply energy transfer
as a continuous evolution of energy storage rather than show it directly
with an arrow. In this specific case, the left-hand bar graph shows
where and how the energy is stored before the person starts pulling,
and the right-hand bar graph shows the energy storage after the pucks
collide. The energy flow diagram shows the transfer of energy from
the person, through the system boundary to the pucks and string. Since
this energy passes through the system boundary, it can be identified
as work.

Teaching with multiple representations allows the teacher access to
parts of students' mental models which would otherwise remain hidden.
Teachers can probe student understanding and get them to think more
deeply by asking questions that make them translate information from
one representation to another. For example, the teacher can ask how
diagrams and graphs should be modified in case certain changes are
made to the problem statement, or how diagrams, graphs and the physical
conditions of a problem would have to change if the equations are
modified. Students also improve their ability to talk to each other
about abstract topics, because debate obtains a concrete focus in
the manipulation of parts of the diagram. This allows a much larger
percentage of the class to follow and therefore participate than if
discussion remained solely in the verbal and mathematical mode. The
use of diagrams, graphs and student-to-student discourse makes it
much more likely that students will think by applying a model and
continually refining it, as opposed to merely memorizing algorithms
or specific answers.\cite{swackhamer-work-work,vanHeuvelen-freshmen-better}

\begin{figure}
\includegraphics{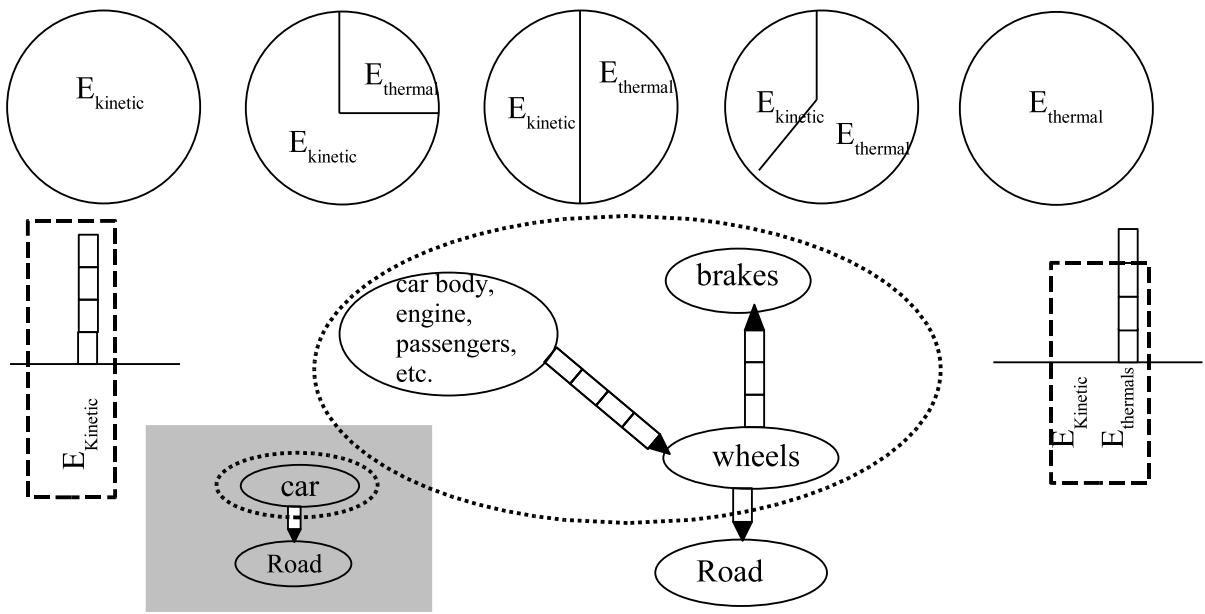}

\caption{\label{fig:diagram-car-kinetic}Energy pie charts, bar graphs and
flow diagram for a car skidding to a stop. The inset shows a simpler
energy flow diagram in which the car is treated as a single object.
Both versions are correct. Which option you choose depends on the
level of detail appropriate for your students and on the concepts
you wish to emphasize.}
\end{figure}

The following problem and variations are discussed by many authors.\cite{arons,swackhamer-work-work,vanHeuvelen-zou}
A car of mass $\mathit{m}$ traveling with velocity $\mathit{V_{o}}$
on a level road travels a distance $\mathit{X}$ while skidding to
a stop. Figure \ref{fig:diagram-car-kinetic} illustrates that the
kinetic energy of the car is transferred to thermal energy stored
in both the car and the road. Only the portion of the kinetic energy
that is actually transferred to the road passes through the system
boundary and can be considered as work. The rest of the thermal energy
initially remains in the car, as represented by the portion of the
bar graph within the dotted system boundary in the right-hand diagram.
Part of this will gradually pass through through the system boundary
as it dissipates, but we have restricted our consideration to an interval
of time before this occurs.\cite{w-e-car-kinetic}

\begin{figure}
\includegraphics{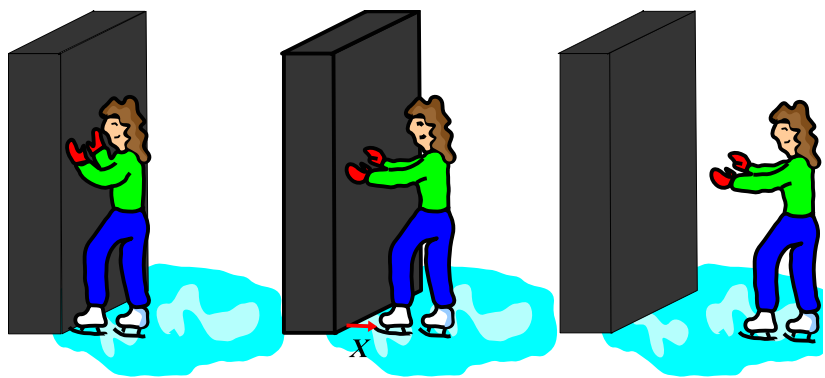}

\caption{\label{fig:skater-wall}A skater stands at rest on frictionless ice
and pushes off of a rigid wall. }
\end{figure}

\begin{figure}
\includegraphics{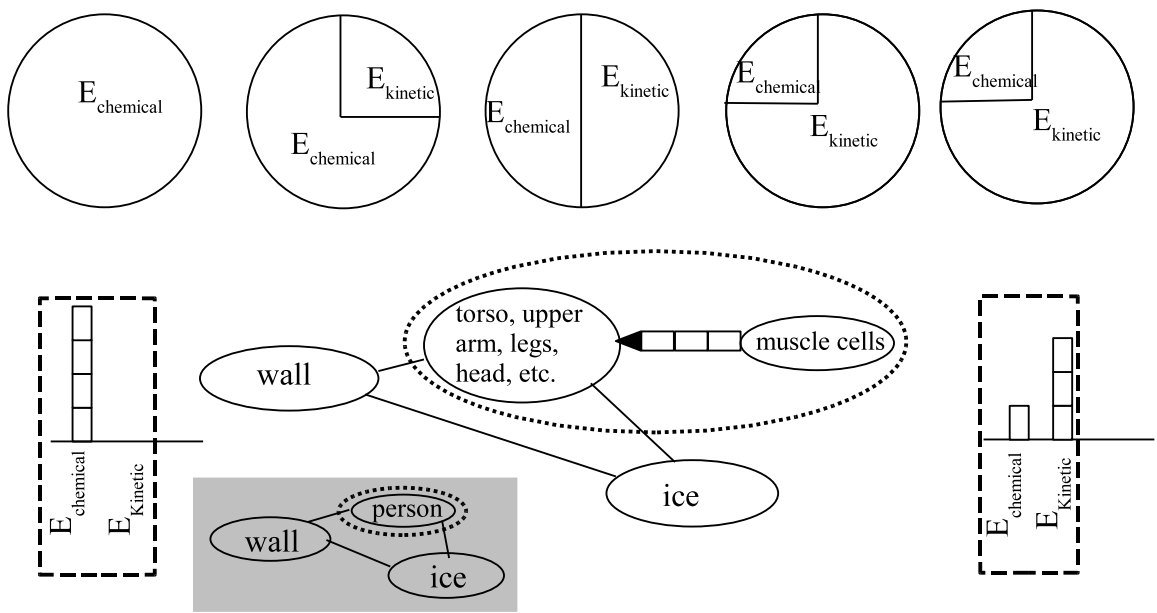}

\caption{\label{fig:wall-skate-diagrams}Energy pie charts, bar graphs and
flow diagram for a skater on frictionless ice pushing off of a rigid
wall. Once again, the inset shows a simpler energy flow diagram in
which the person is treated as a single object. Both versions are
correct. Which option you choose depends on the level of detail appropriate
for your students and on the concepts you wish to emphasize.}
\end{figure}

The following problem involves a so-called {}``zero work force,''
which causes acceleration without doing any work.\cite{arons,mall-leff-stopping,sherwood-pseudowork}
A skater stands at rest on frictionless ice and pushes off of a rigid
wall, as shown in figure \ref{fig:skater-wall}. The skater exerts
a constant horizontal force $\mathit{F}$ and her center of mass is
a distance $\mathit{X}$ farther from the wall after the push. A rigid
wall undergoes no changes as a result of the push, and so does not
change its energy state. Figure \ref{fig:wall-skate-diagrams} shows
that the energy initially stored in the person's muscles ends up stored
as kinetic energy without ever leaving the person. If we choose the
entire person as our system, wishing to concentrate on physics and
avoid the complications of analyzing individual muscle cells in the
person's arm, then the energy transfer in the problem never passes
through the system boundary. Zero work is done in this problem, according
to the only thermodynamically valid definition of work as energy transferred
into our system from external objects.\cite{w-e-wall-skate} Forces
which cause acceleration without transferring energy are actually
quite common. For example, when a person walks, runs, or jumps into
the air, when a car crashes into a brick wall or when a rigid object
rolls without slipping, the wall or floor in each case undergoes no
change and so does not change its energy state. Even though there
is a force between the wall or floor and the system, there is no energy
transfer between these objects and therefore no work.

If students first gain sufficient practice using the diagrams and
graphs to express and refine their ideas, starting with easier problems
and progressing in difficulty, they will continue to use them - and
by implication the conceptual analysis they represent - as the starting
point for quantitative analysis. To help students move from a conceptual-visual
analysis to equations, the curriculum must now provide expressions
for amounts of energy stored and transferred, as well as a way to
associate those expressions with different parts of the diagrams and
graphs, and by implication with the concepts those parts represent.
The overall structure of the diagrams and graphs then allows the student
to write an appropriate conservation of energy equation. In section
III, I will use a two-particle model to derive appropriate expressions
for internal energy and associate them with the visual representations.
In section IV, I simplify those results to four physically sensible
guidelines and show how they can be used to solve the problems we
have just examined conceptually. This theory is mathematically equivalent
to previous theories of internal energy in introductory physics, but
my approach is designed to fit much better with the conceptual and
visual introduction.

\section{The two-particle model}

To apply energy to complex systems, it can be assumed that a macroscopic
object is composed of structureless point particles interacting with
each other by means of conservative, central forces. I will closely
follow the development of multi-body theory in David Hestenes' New
Foundations for Mechanics,\cite{Hestenes-NFCM} but will apply it
only to a two-particle system. The mathematical development is guided
by the idea that the term {}``work'' should receive a clear conceptual
definition as energy that is transferred through the boundary of a
system; a mathematical expression will be identified as work - of
any sort - only when it corresponds to this definition. The two-particle
results can be generalized to a system with an arbitrarily large,
but finite number of particles and the conclusions are similar. 

Our system will be composed of two particles with masses $m_{1}$
and $m_{2}$, positions $\mathbf{X}_{1}$and $\mathbf{X}_{2}$ and
exerting mutual forces $\mathbf{f}_{12}$ and $\mathbf{f}_{21}$,
assumed to obey the third law, to be conservative, directed along
the line joining the two particles, and to depend only on their relative
position, $\mathbf{r}=$$\mathbf{X\mathrm{_{1}-\mathbf{X\mathrm{_{2}}}}}$.
External objects exert a total force $\mathbf{F}_{1}$ on particle
1 and total force $\mathbf{F}_{2}$ on particle 2. In other words:

\begin{eqnarray}
\mathbf{f}_{12}=-\mathbf{f}_{21}=\frac{f_{12}\left(\mathbf{r}\right)\mathbf{r}}{r} & and & \int_{\mathbf{r}_{o}}^{\mathbf{r}_{f}}\mathbf{f}_{12}\cdot d\mathbf{r}=-\Delta U_{int}\label{eq:deltaUint}\\
m_{1}\mathbf{A\mathrm{_{1}=\mathbf{F\mathrm{_{1}+\mathbf{f\mathrm{_{12}}}}}}} &  & m_{2}\mathbf{A\mathrm{_{2}}=}\mathbf{F\mathrm{_{2}+\mathbf{f\mathrm{_{21}=\mathbf{F\mathrm{_{2}-\mathbf{f\mathrm{_{12}}}}}}}}}\label{eq:N2single}\end{eqnarray}
Since these assumptions are properties of almost all fundamental interactions
between elementary particles, the multi-body theory we develop will
have broad applicability. 

Hestenes begins his analysis with the following observation:

\begin{quotation}
To analyze the behavior of a system, we must separate it from its
environment. This is done by distinguishing external and internal
variables. The external variables describe the system as a whole and
its interaction with other (external) systems. The internal variables
describe the (internal) structure of the system and the interactions
among its parts.
\end{quotation}
For the system of both particles, the external variables include:
\begin{eqnarray}
\mathrm{total\: mass} & \qquad & M=m_{1}+m_{2}\label{eq:ext-mass}\\
\mathrm{total\: external\: force} & \qquad & \mathbf{F}=\mathbf{F}_{1}+\mathbf{F}_{2}\label{eq:ext-F}\\
\mathrm{center\: of\: mass} & \qquad & \mathbf{X}=\frac{m_{1}\mathbf{X}_{1}+m_{2}\mathbf{X}_{2}}{M}\label{eq:ext-X}\\
\mathrm{velocity\: of\: center\: of\: mass} & \qquad & \mathbf{V}=\frac{d\mathbf{X}}{dt}=\frac{m_{1}\mathbf{V}_{1}+m_{2}\mathbf{V}_{2}}{M}\label{eq:ext-V}\\
\mathrm{acceleration\: of\: center\: of\: mass} & \qquad & \mathbf{A}=\frac{d\mathbf{V}}{dt}=\frac{m_{1}\mathbf{A}_{1}+m_{2}\mathbf{A}_{2}}{M}\label{eq:ext-A}\\
\mathrm{translational\:(or\: center\: of\: mass)\: kinetic\: energy} & \qquad & E_{kinetic-cm}=\frac{1}{2}MV^{2}\label{eq:ext-KE}\end{eqnarray}

The internal variables describe the motion of the particles relative
to the center of mass as well as their interactions with each other.
These include $m_{1}$, $m_{2}$, $\mathbf{f}_{12}=-\mathbf{f}_{21}$
and $\bigtriangleup U_{int}$ as well as: \begin{eqnarray}
\mathbf{v}_{1}=\mathbf{V}_{1}-\mathbf{V} & \qquad & \mathbf{v}_{2}=\mathbf{V}_{2}-\mathbf{V}\label{eq:int-v}\\
Internal\: Kinetic\: Energy &  & E_{kinetic-int}=\frac{1}{2}m_{1}v_{1}^{2}+\frac{1}{2}m_{2}v_{2}^{2}\label{eq:int-KE}\\
Total\: Internal\: Energy &  & \bigtriangleup E_{int}=\bigtriangleup E_{kinetic-int}+\bigtriangleup U_{int}\label{eq:Eint}\end{eqnarray}
Notice that I use capital letters with subscripts to refer to individual
particle variables, plain capital letters to refer to external variables,
and lower-case letters to refer to internal variables. 

The total internal energy of a multi-particle system is the sum of
the kinetic energy due to motion of the particles relative to the
center of mass and the potential energy stored in the fields between
the particles, which depends on the inter-particle distances. While
it is straightforward to account for the internal energy of a two-particle
system by simply adding these terms, the method quickly becomes unwieldy
as the number of particles in the system increases. For large systems,
it is usually easier to reorganize equation \ref{eq:Eint} into a
smaller number of terms that are more accessible to macroscopic measurements.
Hence, the internal energy includes the thermal energy due to oscillations
of the particles in the system. When the system changes shape because
it is compressed, stretched or bent in any way, then inter-particle
distances must change, altering the internal potential energy. If
such deformations are elastic, then internal kinetic energy may later
change as some part of the system recovers by {}``springing back.''
If the system is capable of undergoing a change in phase or a chemical
change, the resulting changes in inter-particle distances and configurations
will change the internal energy as well.\cite{thermo-mech} Although
it may not be immediately obvious, the internal energy also includes
rotational kinetic energy. Consider the case of a rigid body, for
which the distances of each particle relative to the center of mass
as well as relative to each other are fixed. In that case, the only
way particles can move relative to the center of mass is if the entire
object rotates with angular velocity $\boldsymbol{\omega}$, so that
$E_{kinetic-int}=\frac{1}{2}m_{1}v_{1}^{2}+\frac{1}{2}m_{2}v_{2}^{2}=\frac{1}{2}m_{1}r_{1}^{2}\left(\frac{v_{1}}{r_{1}}\right)^{2}+\frac{1}{2}m_{2}r_{2}^{2}\left(\frac{v_{2}}{r_{2}}\right)^{2}=\frac{1}{2}\left(m_{1}r_{1}^{2}+m_{2}r_{2}^{2}\right)\omega^{2}=\frac{1}{2}I\omega^{2}$,
where $r_{i}=\frac{\left|\mathbf{\boldsymbol{\omega}\times}\mathbf{r}_{i}\right|}{\omega}$
is the distance of the ith particle from the axis of rotation and
I is the moment of inertia about that axis. Introductory texts usually
include a lengthier derivation of rotational kinetic energy and moment
of inertia, and it could be said that I am now applying a similar
treatment to all internal energy.

The total kinetic energy is the sum of the kinetic energy of the two
particles, but it can be separated into internal and external parts
using equations \ref{eq:int-v} , \ref{eq:ext-V} and \ref{eq:ext-mass}.\cite{KE-total-multi}

\begin{multline}
E_{kinetic-total}=\frac{1}{2}m_{1}V_{1}^{2}+\frac{1}{2}m_{2}V_{2}^{2}=\frac{1}{2}m_{1}\left(\mathbf{v}_{1}+\mathbf{V}\right)\cdot\left(\mathbf{v}_{1}+\mathbf{V}\right)+\frac{1}{2}m_{2}\left(\mathbf{v}_{2}+\mathbf{V}\right)\cdot\left(\mathbf{v}_{2}+\mathbf{V}\right)\\
=\frac{1}{2}\left(m_{1}+m_{2}\right)V^{2}+\frac{1}{2}m_{1}v_{1}^{2}+\frac{1}{2}m_{2}v_{2}^{2}+\mathbf{V}\cdot\left(m_{1}\mathbf{v}_{1}+m_{2}\mathbf{v}_{2}\right)=E_{kinetic-cm}+E_{kinetic-int}\label{eq:KE-total}\end{multline}

To relate changes in kinetic energy to the forces on the system, we
take the derivative of \ref{eq:KE-total} and use \ref{eq:N2single}.

\begin{eqnarray}
\frac{dE_{kinetic-total}}{dt} & = & \frac{d}{dt}\left(\frac{1}{2}m_{1}\mathbf{V}_{1}\cdot\mathbf{V}_{1}+\frac{1}{2}m_{2}\mathbf{V}_{2}\cdot\mathbf{V}_{2}\right)=m_{1}\mathbf{V}_{1}\cdot\frac{d\mathbf{V}_{1}}{dt}+m_{2}\mathbf{V}_{2}\cdot\frac{d\mathbf{V}_{2}}{dt}\nonumber \\
 & = & \mathbf{V}_{1}\cdot\left(m_{1}\mathbf{A}_{1}\right)+\mathbf{V}_{2}\cdot\left(m_{2}\mathbf{A}_{2}\right)=\left(\mathbf{F}_{1}+\mathbf{f}_{12}\right)\cdot\mathbf{V}_{1}+\left(\mathbf{F}_{2}+\mathbf{f}_{21}\right)\cdot\mathbf{V}_{2}\label{eq:dKEtot/dt}\\
\frac{dE_{kinetic-total}}{dt} & = & \frac{dE_{kinetic-cm}}{dt}+\frac{dE_{kinetic-int}}{dt}=\left(\mathbf{F}_{1}\cdot\mathbf{V}_{1}+\mathbf{F}_{2}\cdot\mathbf{V}_{2}\right)+\left(\mathbf{f}_{12}\cdot\mathbf{V}_{1}+\mathbf{f}_{21}\cdot\mathbf{V}_{2}\right).\nonumber \end{eqnarray}

The right-hand side of equation \ref{eq:dKEtot/dt} can be expressed
in terms of internal and external variables using equations \ref{eq:int-v}
and \ref{eq:ext-F} .\cite{int-ext-power-multi} \begin{eqnarray}
\mathbf{F}_{1}\cdot\mathbf{V}_{1}+\mathbf{F}_{2}\cdot\mathbf{V}_{2} & \mathbf{=F}_{1}\cdot\left(\mathbf{v}_{1}+\mathbf{V}\right)+\mathbf{F}_{2}\cdot\left(\mathbf{v}_{2}+\mathbf{V}\right) & =\mathbf{F}\cdot\mathbf{V}+\left(\mathbf{F}_{1}\cdot\mathbf{v}_{1}+\mathbf{F}_{2}\cdot\mathbf{v}_{2}\right)\label{eq:ext-power}\\
\mathbf{f}_{12}\cdot\mathbf{V}_{1}+\mathbf{f}_{21}\cdot\mathbf{V}_{2} & =\mathbf{f}_{12}\cdot\left(\mathbf{v}_{1}+\mathbf{V}\right)-\mathbf{f}_{12}\cdot\left(\mathbf{v}_{2}+\mathbf{V}\right) & =\mathbf{f}_{12}\cdot\left(\mathbf{v}_{1}-\mathbf{v}_{2}\right)=\mathbf{f}_{12}\cdot\mathbf{v},\label{eq:int-power}\end{eqnarray}
where the relative velocity is \textbf{$\mathbf{v}=\mathbf{V}_{1}-\mathbf{V}_{2}=\mathbf{v}_{1}-\mathbf{v}_{2}=\frac{d\mathbf{r}}{dt}$}.
Combining equation \ref{eq:dKEtot/dt} with equations \ref{eq:ext-power}
and \ref{eq:int-power} then anti-differentiating and using \ref{eq:deltaUint}
and \ref{eq:Eint}, we find: \begin{equation}
\bigtriangleup E_{kinetic-cm}+\bigtriangleup E_{int}=\bigtriangleup E_{kinetic-cm}+\bigtriangleup E_{kinetic-int}+\bigtriangleup U_{int}=\int_{\mathbf{X}_{0}}^{\mathbf{X}_{f}}\mathbf{F}\cdot d\mathbf{X}+\int_{\mathbf{X}_{10}}^{\mathbf{X}_{1f}}\mathbf{F}_{1}\cdot d\mathbf{x}_{1}+\int_{\mathbf{X}_{20}}^{\mathbf{X}_{2f}}\mathbf{F}_{2}\cdot d\mathbf{x}_{2}\label{eq:dKEext+dEint}\end{equation}

This is the full energy equation for our two-particle system. It presents
us with a sum of three integrals involving dot-products of force and
displacement. In keeping with the previously mentioned guideline,
I defer defining any of these integrals as work until it can be determined
whether they represent energy transferred through the boundary of
the two-particle system. The last two integrals become easier to interpret
if the purely external terms are subtracted from the equation. Start
by differentiating equation \ref{eq:ext-KE} and using \ref{eq:N2single},
\ref{eq:ext-F} and \ref{eq:ext-A} to obtain $\frac{dE_{kinetic-cm}}{dt}=M\mathbf{V}\cdot\frac{d\mathbf{V}}{dt}=\mathbf{V}\cdot\left(M\mathbf{A}\right)=\mathbf{F}\cdot\mathbf{V}$,
which can be anti-differentiated to obtain: \begin{equation}
\bigtriangleup E_{kinetic-cm}=\int_{\mathbf{X}_{0}}^{\mathbf{X}_{f}}\mathbf{F}\cdot d\mathbf{X}\label{eq:work-energy-eqn}\end{equation}
Subtracting equation \ref{eq:work-energy-eqn} from equation \ref{eq:dKEext+dEint},
we find the remaining terms are related by: \begin{equation}
\bigtriangleup E_{int}=\int_{\mathbf{X}_{10}}^{\mathbf{X}_{1f}}\mathbf{F}_{1}\cdot d\mathbf{x}_{1}+\int_{\mathbf{X}_{20}}^{\mathbf{X}_{2f}}\mathbf{F}_{2}\cdot d\mathbf{x}_{2}\label{eq:delta-E-int}\end{equation}
To paraphrase Hestenes\cite{Hestenes-NFCM}, this equation describes
alteration of the internal energy by external forces. Specifically
this occurs when external objects force particles of the system to
move relative to the center of mass. This implies rotation and/or
deformation of the system, both visible consequences of external forces
that can alert students to internal energy transfers as they gradually
build a complete concept of energy. It is precisely when the system
is rotated or deformed that it cannot be approximated as a single
particle and a two-particle model becomes the simplest approach that
retains an accurate understanding of energy. 

This not only provides a meaning for the last two integrals in equation
\ref{eq:dKEext+dEint}, it tells us that the integral in equation
\ref{eq:work-energy-eqn} can be interpreted as the work only when
the the external forces cause no changes in internal energy so that
the two integrals in equation \ref{eq:delta-E-int} sum to zero. In
that case, the only source for the change in center of mass kinetic
energy is the external object or objects exerting the force $\mathbf{F}$.
Of course, one case where this obtains is a single-particle system,
which is the starting point for energy analysis in most introductory
texts. While equation \ref{eq:work-energy-eqn} says literally that
the integral of the dot product of force and center of mass displacement
equals a change in center of mass (translational) kinetic energy,
many texts proceed to adopt $\int_{\mathbf{X}_{0}}^{\mathbf{X}_{f}}\mathbf{F}\cdot d\mathbf{X}$as
a general definition of work and apply this definition outside its
limited range of validity. As section II shows, situations where external
forces cause changes in internal energy are quite common and are almost
always included among the examples and problems in these same introductory
texts. Conceptual confusion sets in when $\int_{\mathbf{X}_{0}}^{\mathbf{X}_{f}}\mathbf{F}\cdot d\mathbf{X}$
is called {}``work'' but does not represent energy transfer through
the system boundary. It is not equation \ref{eq:work-energy-eqn}
itself which is the problem, but its interpretation as a general definition
of work which leads to conceptual difficulties. Other authors \cite{arons,arons-book,Mungan-review,sherwood-pseudowork,swackhamer-work-work,mall-leff-all-about-work}recognize
the limits of the definition and supplement equation \ref{eq:work-energy-eqn}
with a multi-particle energy equation while inventing names such as
{}``particle work'' or {}``pseudo work'' to identify$\int_{\mathbf{X}_{0}}^{\mathbf{X}_{f}}\mathbf{F}\cdot d\mathbf{X}$
in troublesome situations. As an alternative, I continue the mathematical
development without applying names until I can clearly identify which
integrals truly represent energy transferred through the system boundary. 

Our investigation of two-particle models has produced two different
types of force vs. distance integrals, which differ in the choice
of external or internal variables for the position. For example, $\int_{\mathbf{X}_{10}}^{\mathbf{X}_{1f}}\mathbf{F}_{1}\cdot d\mathbf{x}_{1}$
can be used to calculate an amount of energy transferred from external
objects to particle 1 and stored internally by the system. On the
other hand, $\int_{\mathbf{X}_{10}}^{\mathbf{X}_{1f}}\mathbf{F}_{1}\cdot d\mathbf{X}$
(which is part of $\int_{\mathbf{X}_{0}}^{\mathbf{X}_{f}}\mathbf{F}\cdot d\mathbf{X}$
because of equation \ref{eq:ext-F}) represents energy transferred
from external objects to particle 1 and stored as translational kinetic
energy of the system. In addition, these two integrals can be added
to get: \begin{equation}
\int_{\mathbf{X}_{10}}^{\mathbf{X}_{1f}}\mathbf{F}_{1}\cdot d\mathbf{x}_{1}+\int_{\mathbf{X}_{10}}^{\mathbf{X}_{1f}}\mathbf{F}_{1}\cdot d\mathbf{X}=\int_{\mathbf{X}_{10}}^{\mathbf{X}_{1f}}\mathbf{F}_{1}\cdot d\left(\mathbf{x}_{1}+\mathbf{X}\right)=\int_{\mathbf{X}_{10}}^{\mathbf{X}_{1f}}\mathbf{F}_{1}\cdot d\mathbf{X}_{1}\label{eq:F-dot-X-total-particle}\end{equation}
The right-hand side of this equation involves the total force from
external objects on particle 1 as well as the total displacement of
particle 1. This is the total amount of energy transferred from external
objects to particle 1, and thus to the system. To be consistent with
the thermodynamic definition of work as the energy transferred through
the system boundary, this must be the work done by forces $\mathbf{F}_{1}$
on the system. The left side of equation \ref{eq:F-dot-X-total-particle}
(along with equations \ref{eq:work-energy-eqn} and \ref{eq:dKEext+dEint})
tells us that this total energy transfer accounts for a portion of
the changes in internal energy and translational kinetic energy. While
this may seem abstract, when solving elementary problems it is often
possible to model the system as two particles, one particle located
at the point of application of the force and another located at the
center of mass and representing the rest of the object. From there,
one can identify displacement of the center of mass, displacement
of the first particle relative to the center of mass and total displacement
of the particle. Even when one or more of these displacements is unknowable,
distinguishing among the three cases can aid an analysis of energy
storage and transfer and help resolve confusion.

The interpretation of force vs. displacement integrals in this paper,
in particular the decision to apply the term work in general to equation
\ref{eq:F-dot-X-total-particle} but not equations \ref{eq:work-energy-eqn}
or \ref{eq:delta-E-int}, is a direct consequence of adopting an appropriate
mathematical model - one capable of representing internal energy because
the system includes more than one particle - and strictly defining
{}``work'' in accord with thermodynamics, as the amount of energy
transferred from external objects through the boundary of the system.
Mallinckrodt and Leff\cite{mall-leff-all-about-work} define seven
different force vs. displacement integrals that appear in the literature,
but point out that only three of these are independent. Starting with
any three, the other four may be determined by simple mathematical
relationships. My choice of three integrals is therefore a sufficient
set and mathematically equivalent to prior formulations. Mallinckrodt
and Leff generally regard this choice as the easiest set of three
to understand and apply. 

Finally, it should be noted that when external objects exert conservative
forces on the multi-particle system, the system may be enlarged to
include these objects and introduce external potential energy terms
to the left-hand side of equation \ref{eq:dKEext+dEint}. However,
changes to the internal energy may also occur if the forces in question
are not uniform. Tidal forces are one example of this, and they work
precisely because non-uniform gravitational forces deform an object
by displacing parts of it relative to the center of mass. Use of two-particle
models for elementary problems can help prepare students for the study
of more advanced topics, such as tidal forces, when they encounter
them later in their studies. In many cases, the external conservative
force is approximately uniform and will therefore cause only negligible
changes in internal energy.

\section{Solving problems by combining diagrams, graphs and mathematics}

Zou and vanHeuvelen outline a general problem-solving procedure when
employing diagrams and graphs.\cite{vanHeuvelen-zou} The student
first analyzes energy storage and transfer using the pie charts, bar
graphs, and energy flow diagrams as in section II. In the case of
a deformable or rotating system, the guidelines developed in section
III help the student associate each energy formula with the appropriate
energy transfer or storage mechanism. There are three displacements
to examine for each external force: the displacement $\mathbf{X}$
of the center of mass, the displacement $\mathbf{x}$ - relative to
the center of mass - of the particle to which the force is directly
applied, and the total displacement $\mathbf{X}_{1}=\mathbf{x}+\mathbf{X}$
of that particle. If the force is constant, $FX\cos\theta$ will equal
the change in translational kinetic energy; $Fx\cos\theta$ equals
the change in internal energy; and $FX_{1}\cos\theta$ equals the
total energy transferred from the external object to the system, which
is the work. For basic mechanics, the kinetic energy formula is also
necessary. Others, such as the specific heat formula, can be added
as the course sequence dictates. The student associates each energy
formula with one or more parts of the diagrams, and reads a conservation
of energy equation from the overall structure. \textit{The only necessary
mathematical change to the curriculum is to provide students with
the meaning of the above three products in lieu of defining work as
$FX\cos\theta$.} The mathematical manipulations achieved by this
procedure are identical to those recommended by previous authors.
The single particle energy model is discarded, and students learn
a multi-particle theory first conceptually and then mathematically.
When appropriate, the conservation of energy equation obtained by
this procedure will automatically simplify to the single-particle
work-energy equation.

\begin{figure}
\includegraphics{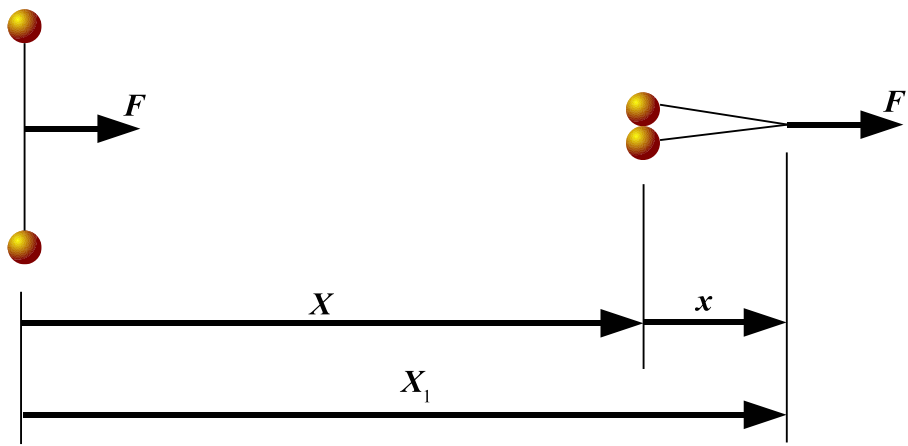}

\caption{\label{fig:2pucks-int-ext}Internal, external and total particle
displacements for the two-puck-and-string problem.}
\end{figure}

\begin{figure}
\includegraphics{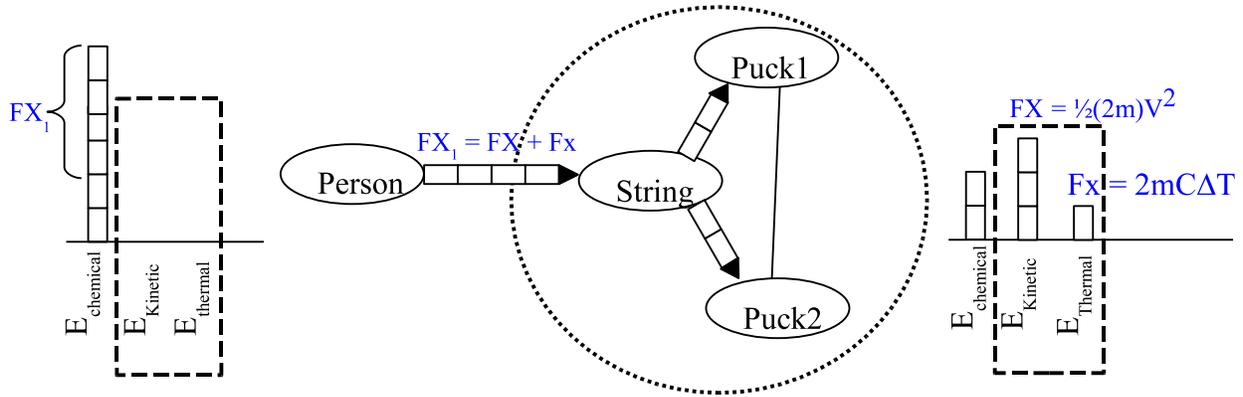}

\caption{\label{fig:2pucks-eqn-diagram}Mathematical expressions and equations
have been added to the energy bar graphs and flow diagram from figure
3 to show how each corresponds to one or more parts of the diagram.
The structure of the diagrams and graphs leads directly to a conservation
of energy equation.}
\end{figure}

Figure \ref{fig:2pucks-int-ext} is a more complete diagram for the
two-pucks-and-string problem from section II, which reveals that the
point of application of the force (the center of the string) has been
displaced relative to the center of mass, causing the system to change
shape. Hence, we know that $FX$ will not equal the work. Instead,
figure \ref{fig:2pucks-eqn-diagram} shows how to correctly associate
energy formulas with parts of the diagrams. $FX=\frac{1}{2}\left(2m\right)V^{2}$,
because products with the center of mass displacement equal changes
in translational kinetic energy. $Fx=F\left(X_{1}-X\right)=2mC\Delta T$
(where C is the specific heat and T is the temperature) because products
with the internal displacement equal the changes in internal energy,
which is thermal in this case. Finally, $FX_{1}=\frac{1}{2}\left(2m\right)V^{2}+2mC\Delta T$,
because the total work done by the person is transferred to the pucks
and stored both internally and externally. There is a clear conceptual
interpretation for each product of force and displacement, and it
is clear that $FX_{1}$ is the work, while $FX$ and $Fx$ are not.\cite{2pucks-confirm}
A student who can draw correct diagrams, explain the meaning of those
diagrams in words, and associate each part of their calculation with
part of a diagram has demonstrated a thorough understanding both conceptually
and mathematically. The mathematics is not only consistent with the
conceptual analysis, it follows directly from the concepts in a straightforward
manner. A larger portion of the energy unit may be spent on conceptual
analysis, because the structure of the diagrams and graphs leads so
neatly into the conservation of energy equation. Students are far
less likely to make common mathematical mistakes such as negative
sign errors or double accounting for work and potential energy, so
less time is required to address mathematical difficulties.

\begin{figure}
\includegraphics{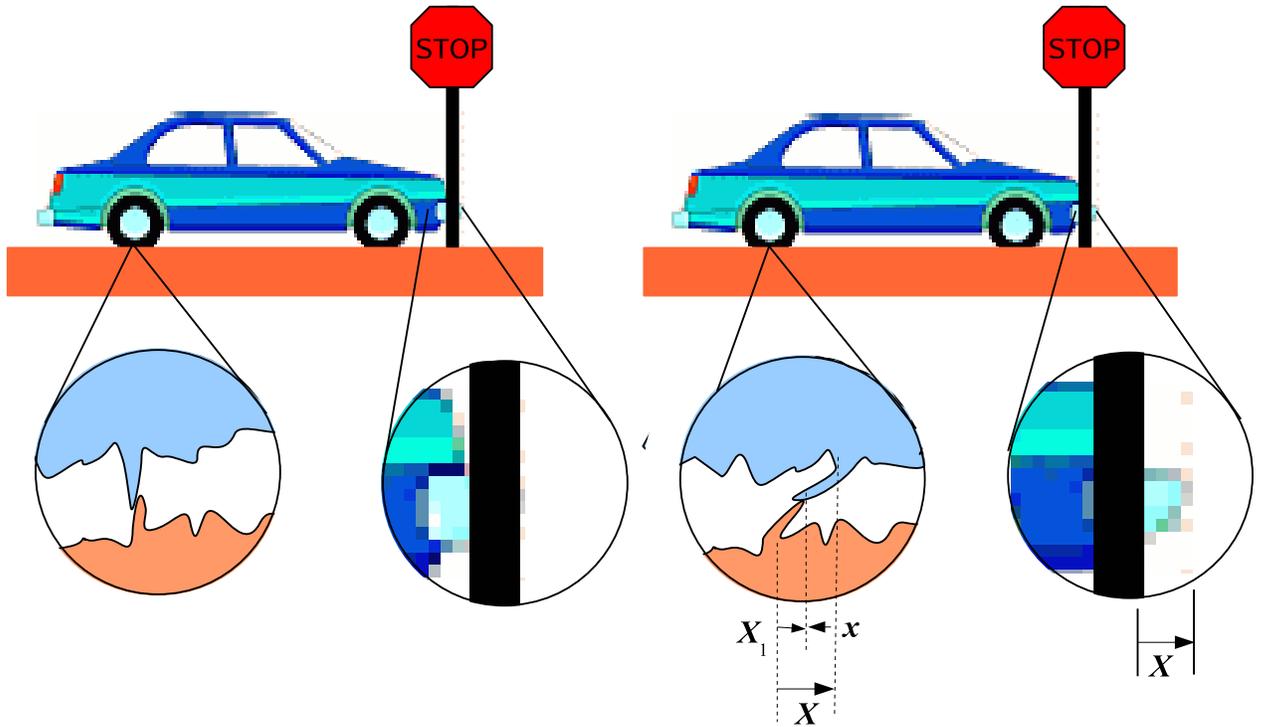}

\caption{\label{fig:car-skid-detail}Internal and External Variables for an
Object Skidding to a Stop. The car starts to skid at the instant its
front end reaches the pole of the stop sign. The second and fourth
blowups show the car moving a very small (actually microscopic) distance
$\mathbf{X}$ during the first few instants of the skid. The first
and third blowups are inspired by Sherwood and Bernard's {}``stylized
microscopic view of friction, with greatly exaggerated vertical scale,''
showing that {}``teeth'' belonging to the tire and the road surface
have welded together at their contact point.\cite{sherwood-heat-work}
The upper tooth is carried forward a distance $\mathbf{X}$. The point
where the force is applied at the tip of the tooth is displaced a
distance $\mathbf{x}$ to the left relative to the center of mass,
but relative to the lower tooth, it is displaced a distance  $\mathbf{X}_{1}=\mathbf{X}-\mathbf{x}$
to the right. Scales in this figure have been exaggerated disproportionately
for the sake of clear presentation of the relation among the displacement
vectors.}
\end{figure}

The frictional force on a car, or any object, skidding to a stop is
actually the sum of many microscopic forces between the surface (the
road) and the skidding object (the car).\cite{sherwood-heat-work}
As the car skids forward, a {}``tooth'' from the uneven road surface
exerts an external force $\mathbf{f}$ on a tooth of the car tire
opposite to the direction of travel (see figure \ref{fig:car-skid-detail}).
This causes a negative change in the center of mass kinetic energy
equal to $-fX$, because the frictional force is opposite the displacement
of the center of mass, $\mathbf{X}$. It causes a positive change
in the internal energy equal to $fx$, because the frictional force
and the internal displacement of the tire atom, $x$ are in the same
direction, both opposite the motion. And there is a net transfer of
energy out of the system equal to $-f\left(X-x\right)=-fX_{1}$. $-fX_{1}$
is the work, but we are unable to calculate an exact value without
knowing the microscopic displacement $x$. The total energy transfer
will be the sum of similar terms over all of the atoms of the tire
in contact with atoms of the road. While we cannot calculate the work,
if the total external frictional force and the displacement of the
center of mass are known, we can calculate the total change in the
translational kinetic energy of the center of mass. This is the calculation
typically accepted as a solution to this problem, and it is mathematically
correct. However, it is conceptually misleading to claim that it expresses
the idea that work done by the frictional force equals the change
in the kinetic energy. It does not. \cite{negative-signs}

\begin{figure}
\includegraphics{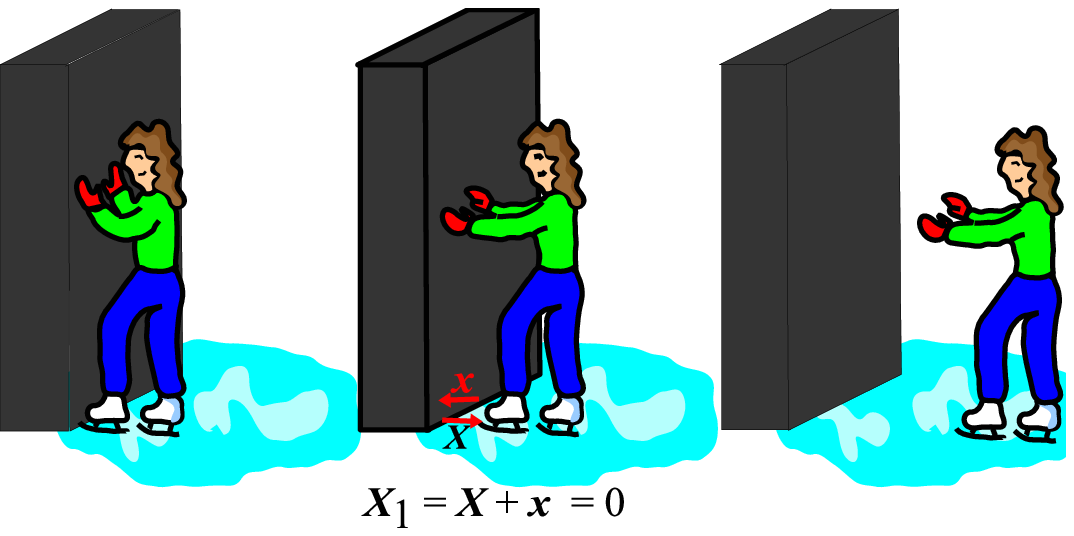}

\caption{\label{fig:skate-wall-detail}Internal and external variables for
a skater on frictionless ice pushing off of a wall. The person's center
of mass was displaced a distance $\mathbf{X}$ during the push, while
the point of application of the force was displaced an equal amount
in the opposite direction, relative to the center of mass, but was
not displaced at all relative to the wall. }
\end{figure}

Recall the skater pushing off of the wall? The diagrams in section
II showed that no work was done because no energy was transferred
from the wall to the person. The person changes shape or is {}``deformed''
when the point of application of the force, the person's hands, is
displaced relative to the center of mass. Therefore, we know that
$FX$ will not equal the work. In more detail, figure \ref{fig:skate-wall-detail}
shows that the person's center of mass was displaced a distance $\mathbf{X}$
during the push, while the point of application of the force was displaced
an equal amount in the opposite direction, relative to the center
of mass, but was not displaced at all relative to the wall. $FX$
represents an increase in the translational kinetic energy of the
person. $-Fx$ represents an equal decrease in the internal energy
of the person, corresponding exactly with our conceptual analysis.
Finally $FX_{1}$ is the work, the amount of energy transferred from
the wall to the person, and it is indeed zero.

\begin{figure}
\includegraphics{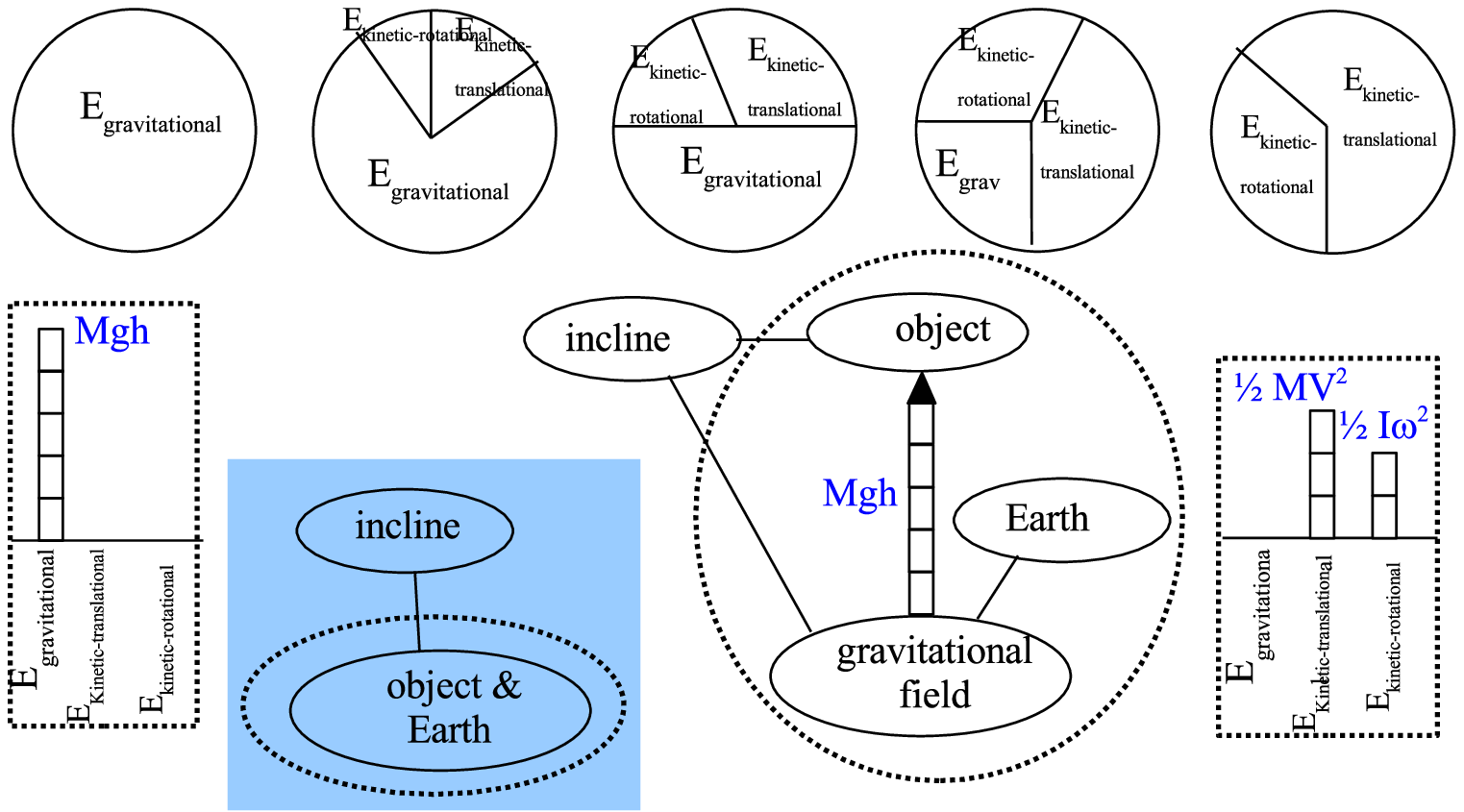}

\caption{\label{fig:roll-diagrams}Energy pie charts, bar graphs and flow
diagram for a round object rolling down an incline. The inset shows
a simpler energy flow diagram in which the round object, the Earth
and by implication also the gravitational field are treated as a single
object. Both versions are correct. Which option you choose depends
on the level of detail appropriate for your students, in particular
their readiness to deal with the concept of the field as real physical
object which, although invisible, stores the gravitational potential
energy.\cite{sherwood-gravity}}
\end{figure}

\begin{figure}
\includegraphics{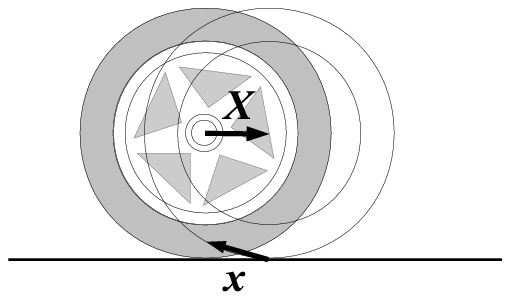}

\caption{\label{fig:roll-detail}Infinitesimal Internal and External Displacements
for a Rolling Object - In an infinitesimal time interval $\Delta t$,
the center of mass of a rolling object is displaced by an amount $\mathbf{X}$,
and the original point of contact is displaced by an amount $\mathbf{\mathbf{x}}$
relative to the center of mass. If the object rolls without slipping,
$\mathbf{\mathbf{x}}$ is approximately equal in magnitude but opposite
in direction to $\mathbf{X}$, so that in the limit that $\Delta t$
approaches zero, the displacement of the point of contact is also
zero.}
\end{figure}

As a final example, consider a round object of mass M, radius R and
moment of inertia $I=bMR^{2}$, where b is a unitless fraction equal
to $\frac{2}{5}$ for a sphere or $\frac{1}{2}$ for a cylinder, etc.
When the object starts from rest and rolls without slipping down an
incline of height $H$ and length $L$, gravitational energy is transferred
to translational kinetic energy plus rotational kinetic energy, as
shown in figure \ref{fig:roll-diagrams}. This gives $MgH=\frac{1}{2}MV^{2}+\frac{1}{2}I\omega^{2}=\frac{1}{2}MV^{2}\left(1+b\right),$where
$V=\omega R$ because the object does not slip. The static frictional
force, $f$ acts opposite the direction of motion while the object's
center of mass is displaced a distance $L$ down the incline. The
meaning of the product $fL$ can be determined by examining infinitesimal
displacements of a rolling object, as shown in figure \ref{fig:roll-detail}.
Because the point of application of the frictional force (where the
object contacts the ground) is displaced relative to the center of
mass, we can anticipate that the dot product of the frictional force
and the displacement of the center of mass will not equal the work.
The frictional force $f$ is opposite in direction to the displacement
of the center of mass $X$, leading to a decrease in the object's
translational kinetic energy, $-fX=\Delta\left(\frac{1}{2}MV^{2}\right)$.
On the other hand, $f$ is in approximately the same direction as
the internal displacement, $x$, leading to an increase in the object's
internal energy, $fx=\Delta\left(\frac{1}{2}I\omega^{2}\right)$.
If the object rolls without slipping, then in the limit $\Delta t$
approaches zero, $\mathbf{x}$ is equal and opposite to $\mathbf{X}$,
so that the instantaneous displacement of the point of contact is
$\mathbf{X\mathrm{_{1}}=}\mathbf{X-x\mathrm{=0}}$ and no work is
done. The action of the frictional force is to transfer energy from
translational kinetic energy to rotational kinetic energy without
doing any work because the energy remains in the system at all times.\cite{w-e-roll}
Hence, there is no energy flow arrow betwen the incline and the object
in figure \ref{fig:roll-diagrams}, and we can conclude that $fL=\Delta KE_{rotational}=\frac{bMgH}{b+1}=\frac{1}{2}I\omega^{2}$.
Carnero et. al, Sherwood and Mungan obtain the same result,\cite{carnero-roll-work,Mungan-review,sherwood-pseudowork}
but the meaning of the force-displacement products derived from the
two particle model provides a much more straightforward path to this
solution. Although one cannot strictly model a rolling object with
a two-particle model, the introduction of two-particle models earlier
in the course sequence prepares students for a more sophisticated
understanding of work and energy in rolling motion, as a comparison
of the pairs of equal and opposite displacements in figures \ref{fig:skate-wall-detail}
and \ref{fig:roll-detail} shows.

\section{conclusion}

One goal of a good introductory physics course should be for students
to develop a useful model of energy and its conservation, one that
they can employ to understand energy concepts throughout their lives.
Teachers must take care to present the material in a logical sequence
so that concepts are continually refined and ideas build on one another.
At the introductory level, the mathematics should be kept as simple
as possible, but a certain level of complexity is necessary in order
to support the development of correct energy concepts. In this conclusion,
I will outline some successful reforms of the pedagogical sequence
for teaching energy, point out where two-particle models fit into
this sequence, and show how the inclusion of two-particle models for
some elementary problems can help prepare students for more advanced
topics.

The overarching theme of Eric Brewe's doctoral disseration\cite{brewe}
is {}``energy early, energy often, energy intuition.'' He spends
approximately the same amount of class time on energy as he would
in a more standard course sequence, but instead of covering energy
all at once, he weaves in an energy thread throughout. During the
first week, he teaches students to represent their energy ideas qualitatively
using pie charts and system schemas.\cite{system-schema} At some
point during each unit, he revisits energy by applying these representational
tools to new situations, while gradually refining the existing tools
and adding others - energy flow arrows on system schemas, bar graphs,
equations and potential graphs. Students do not use energy equations
until they have developed a more sophisticated understanding of energy
conservation. It is important that students practice using the diagrams
and graphs to express energy ideas for some time before proceeding
to equations. If they are comfortable with the representational tools,
they will continue to use them as an aid to problem-solving. The result
of Brewe's reform is greater comfort with energy as a problem-solving
alternative and a greater understanding of its relation to other topics.\cite{extension-to-modeling}

Within Brewe's sequence, two-particle models should not be introduced
until students understand how to apply the representational tools
and equations to simpler situations. These include situations in which
internal energy changes are negligible, as well as other cases, such
as many inelastic or partially elastic collisions, where modeling
objects as single particles that can store internal energy may be
sufficient. In the two puck and string problem, for instance, we did
just that when we assumed that a portion of the pucks' kinetic energy
was transferred to thermal energy when they collided. Once students
have an appropriate background, examples such as those from section
II can be used to help them refine their definition of work and to
develop a deeper understanding of energy storage and transfer. \textit{Because
the conceptual development has included internal energy from the start
and the structure of the diagrams leads directly to the equations,
one need only introduce the meaning of the three different force vs.
displacement products to apply the two-particle model. }

A common bouncing ball occupies a pivotal point in the calculus-based
college physics course as taught by Dwayne Desbian, and shows how
students can be led to understand the difference between a single-particle
and a multi-particle model.\cite{desbain} This situation contains
basic elements of almost all of the examples in this paper. The floor
causes the ball to flex as it comes momentarily to rest, in a manner
similar to the way the brick wall crumples the car's front end. As
the ball rebounds, some of the energy is restored to kinetic energy
because of the elastic flexing of the ball's material. This energy
does not come from the floor, but in a manner similar to the jumping
person or the skater pushing off from the wall, it remains stored
in the ball the entire time. Even though the ball may store less kinetic
energy after the bounce than before, that does not mean the energy
has left the ball. Similar to the car skidding to a stop, the energy
is at least partially stored as thermal energy in the ball itself.
Through careful consideration of this problem using all of the different
representations, students in Desbian's course discover the failure
of the single particle model and develop a model of the ball as two
particles mediated by a linear restoring force. 

I showed in section IV how an early introduction to internal energy
better prepares students to understand energy transfer when objects
roll without slipping. Students who continue their physics education
will be exposed to ideas of internal energy and deformation in even
more complicated contexts, including thermodynamics, molecular dynamics,
solid-state physics, and atomic physics.\cite{alonso-finn2} In thermodynamics,
students encounter difficulty understanding the work done on or by
the working fluid of a heat engine. Problems such as the skater and
the car crash provide more concrete and familiar contexts in which
to associate compression and expansion of the system with changes
in internal energy. The two-particle, linear-restoring-force model
of the bouncing ball is important in its own right. Although seemingly
simple, it contains the basic ideas of elastic behavior, and internal
oscillations leading to thermal energy.\cite{mungan-2blocks-spring}
In their introductory physics course, \emph{Matter and Interactions},
Ruth Chabay and Bruce Sherwood make innovative use of particles connected
by linear restoring forces to demonstrate how bulk properties of matter
emerge by applying basic mechanics to its constituent parts.\cite{m&i}
This is the central concept of solid state physics. In atomic physics,
absorption or emission of a photon leads to changes in the internal
energy of an atom, accompanied by deformation in the form of changes
in its size and shape. Once students express atomic transitions using
the familiar bar graphs and energy flow diagrams, they find energy
level diagrams much easier to comprehend. Tides and other phenomena
due to the action of non-uniform fields can be understood through
internal energy transfers accompanied by deformation. Introducing
two-particle models at the introductory level also exposes students
to the idea of a model and its limits, and helps them to understand
multi-body theory when they encounter it in an advanced physics course.

Teaching energy using multiple representations to promote student-to-student
dialog is gaining currency among physics teachers. This paper is but
the latest in a long series of efforts to deal with the mathematics
of internal energy at the introductory level. Both mathematical and
visual approaches are finding their way into current textbooks. On
the other hand, a unitary fluid-like metaphor for energy, a more rigorous
expression of conservation of energy, an elevation of the importance
of diagrams and graphs relative to equations, or the idea that $\int\mathbf{F}\cdot d\mathbf{X}$
does not, in general, equal work are counter-intuitive ideas to many
trained physicists and introductory physics teachers. When they encounter
these ideas in the PER literature, it may well cause them to disregard
the reform efforts. It is my hope that combining the mathematical
and visual approaches and clarifying the mathematical and theoretical
basis for these reforms will help some physicists and teachers take
a closer look at the real value of the proposed changes.

\begin{description}
\item [{Matt~Greenwolfe}] grew up in Indiana and received a B.S. degree
in physics from Washington University in St. Louis and a PhD in physics
from The University of Michigan. Over the past eleven years his high
school classes have evolved to use guided inquiry, student discourse,
and multiple representations. He is currently president of the American
Modeling Teachers Association, www.modelingteachers.org. 
\end{description}
\begin{acknowledgments}
I would like to acknowledge helpful conversations with Dan Yaverbaum,
Liz Quinn-Stein, Dick Mentock, members of the ASU advanced modeling
class from summer 2005, and participants on the modeling physics list
serve. I would also like to thank my wife Joy Greenwolfe for editorial
and graphic design assistance , Carol Hamilton for editorial assistance,
helpful suggestions from David Hestenes that significantly strengthened
this paper, my modeling workshop leaders Art Woodruff and Patty Blanton,
participants in modeling workshops I have taken and led, and the faculty
and staff of Cary Academy for supporting my teaching over the past
six years.
\end{acknowledgments}


\begin{thebibliography}{10}
\bibitem{work-energy-only-dyn}Although the use of the work-energy
theorem is also usually accompanied with a warning to the effect that
it does not give valid information about work and energy, but is merely
a useful dynamical relationship among the variables.
\bibitem{w-e-2puck}The single-particle work-energy theorem shows
that $FX=\frac{1}{2}MV^{2}=\Delta E_{kinetic}$. FX accounts for only
the kinetic energy, but not the thermal energy, and therefore does
not account for all of the energy transferred from the person to the
system. FX is therefore not equal to the work in this situation.
\bibitem{w-e-car-kinetic}The single-particle work-energy theorem
shows that $-fX=-\frac{1}{2}MV_{0}^{2}=\Delta E_{kinetic}$, where
f is the kinetic frictional force between the road and the wheels.
But -fX does not equal the work, because it does not account for the
thermal energy remaining in the car.
\bibitem{w-e-wall-skate}The single-particle work-energy theorem shows
that $FX=\frac{1}{2}MV^{2}=\Delta E_{kinetic}$, but if no work is
done, this formula cannot be described as work equaling the change
in kinetic energy.
\bibitem{sherwood-gravity}As Sherwood\cite{sherwood-pseudowork}
states, {}``It is unfortunately very common to speak of the gravitational
potential energy of a person or of a falling rock, but such statements
really should be avoided since the Earth must be included in the system.''
See also references \cite{arons,falk-herrmann,vanHeuvelen-zou}.
\bibitem{w-e-roll}I would also argue that since fL is not work, it
should also not be labeled {}``pseudo-work,'' {}``particle-work,''
or {}``center-of-mass-work.'' Such semantic distinctions are likely
to confuse an introductory student. On the other hand, student understanding
is readily apparent from their attempts to draw the diagrams. If a
student can trace where the energy starts, where it is transferred,
and where it ends up, then they have a sufficient understanding regardless
of the terminology.
\bibitem{gunter-multi-rep}van Heuvelen and Zou\cite{vanHeuvelen-zou}
quote Howard Gunter, {}``Genuine understanding is most likely to
emerge ... if people possess a number of ways of representing knowledge
of a concept or skill and can move readily back and forth among these
different ways of knowing.''
\bibitem{cm-multi}With a total of N particles, we get: $m_{i}\mathbf{A}_{i}=\mathbf{F}_{i}+\sum_{j\neq i}\mathbf{f}_{ij}$,
$\mathbf{X}=\frac{\sum_{i}m_{i}\mathbf{\mathbf{X}}_{i}}{M}$, and
$M\mathbf{A}=\sum_{i}m_{i}\mathbf{A}_{i}=\sum_{i}\mathbf{F}_{i}+\sum_{i}\sum_{j\neq i}\mathbf{f}_{ij}=\mathbf{F}+\frac{1}{2}\left(\sum_{i}\sum_{j\neq i}\mathbf{f}_{ij}+\sum_{j}\sum_{i\neq j}\mathbf{f}_{ji}\right)=\mathbf{F}+\frac{1}{2}\sum_{i}\sum_{j\neq i}\left(\mathbf{f}_{ij}+\mathbf{f}_{ji}\right)=\mathbf{F}.$
\bibitem{KE-total-multi}Notice that $m_{1}\mathbf{v\mathrm{_{1}+m_{2}\mathbf{v\mathrm{_{2}=m_{1}(\mathbf{V\mathrm{_{1}-\mathbf{V\mathrm{)}-\mathrm{m_{2}(}}\mathbf{V\mathrm{_{2}-\mathbf{V\mathrm{)=m_{1}\mathbf{V\mathrm{_{1}+m_{2}\mathbf{V\mathrm{_{2}-(m_{1}+m_{2})\mathbf{V}}}}}}}}}}}}}}}$
$=\mathrm{M\mathbf{V-\mathrm{M\mathbf{V}=}\mathrm{0.}}}$ With N particles:
$E_{kinetic-total}=\sum_{i}\frac{1}{2}m_{i}V_{i}^{2}$ = $\sum_{i}\frac{1}{2}m_{i}\left(\mathbf{v}_{i}+\mathbf{V}\right)\cdot\left(\mathbf{v}_{i}+\mathbf{V}\right)$
=$\sum_{i}\frac{1}{2}m_{i}V^{2}+\sum_{i}\frac{1}{2}m_{i}v_{i}^{2}+\mathbf{V}\cdot\sum_{i}m_{i}\mathbf{v}_{i}$
= $\frac{1}{2}MV^{2}+E_{kinetic-int}+\mathbf{V}\cdot\sum_{i}m_{i}\left(\mathbf{V}_{i}-\mathbf{V}\right)$
= $\frac{1}{2}MV^{2}+E_{kinetic-int}+\mathbf{V}\cdot\left(\sum_{i}m_{i}\mathbf{V}_{i}-M\mathbf{V}\right)=E_{kinetic-cm}+E_{kinetic-int}$.
\bibitem{int-ext-power-multi}With N particles, \ref{eq:ext-power}
is straightforward, $\sum_{i}\mathbf{F}_{i}\cdot\mathbf{V}_{i}=\sum_{i}\mathbf{F}_{i}\cdot\left(\mathbf{v}_{i}+\mathbf{V}\right)=\left(\sum_{i}\mathbf{F}_{i}\right)\cdot\mathbf{V+\sum_{i}\mathbf{F}_{i}\cdot\mathbf{v}_{i}}=\mathbf{F}\cdot\mathbf{V+}\mathbf{\sum_{i}\mathbf{F}_{i}\cdot\mathbf{v}_{i}}.$
But \ref{eq:int-power} is trickier, $\sum_{i}\sum_{j\neq i}\mathbf{f}_{ij}\cdot\mathbf{V}_{i}=\frac{1}{2}\left(\sum_{i}\sum_{j\neq i}\mathbf{f}_{ij}\cdot\mathbf{V}_{i}+\sum_{j}\sum_{i\neq j}\mathbf{f}_{ji}\cdot\mathbf{V}_{j}\right)=\frac{1}{2}\sum_{i}\sum_{j\neq i}\left(\mathbf{f}_{ij}\cdot\left(\mathbf{v}_{i}+\mathbf{V}\right)-\mathbf{f}_{ij}\cdot\left(\mathbf{v}_{j}+\mathbf{V}\right)\right)$$=\frac{1}{2}\sum_{i}\sum_{j\neq i}\mathbf{f}_{ij}\cdot\left(\mathbf{v}_{i}-\mathbf{v}_{j}\right)=\frac{1}{2}\sum_{i}\sum_{j<i}\mathbf{f}_{ij}\cdot\left(\mathbf{v}_{i}-\mathbf{v}_{j}\right)+\frac{1}{2}\sum_{i}\sum_{j>i}\mathbf{f}_{ij}\cdot\left(\mathbf{v}_{i}-\mathbf{v}_{j}\right)$$=\frac{1}{2}\sum_{i}\sum_{j<i}\mathbf{f}_{ij}\cdot\left(\mathbf{v}_{i}-\mathbf{v}_{j}\right)+\frac{1}{2}\sum_{j}\sum_{i>j}\mathbf{f}_{ji}\cdot\left(\mathbf{v}_{j}-\mathbf{v}_{i}\right)=\frac{1}{2}\sum_{i}\sum_{j<i}\mathbf{f}_{ij}\cdot\left(\mathbf{v}_{i}-\mathbf{v}_{j}\right)+\frac{1}{2}\sum_{i}\sum_{j<i}\mathbf{f}_{ij}\cdot\left(\mathbf{v}_{i}-\mathbf{v}_{j}\right)$$=\sum_{i}\sum_{j<i}\mathbf{f}_{ij}\cdot\left(\mathbf{v}_{i}-\mathbf{v}_{j}\right)=\sum_{i}\sum_{j<i}\mathbf{f}_{ij}\cdot\mathbf{v}_{ij}$,
where $\mathbf{v}_{ij}$is the relative velocity of particles i and
j. In step 5, I re-wrote the second term swapping the indices $i$
and $j$. In the sixth step, I switched the order of the sum and used
Newton's Third Law. Notice that in the final sum, each pair of particles
is counted only once.
\bibitem{still-three-integrals}In the case of more than two particles,
there are still only three types of force vs. displacement integral.
See footnote \cite{int-ext-power-multi} as well as references \cite{Hestenes-NFCM}
and \cite{mall-leff-all-about-work} for details.
\bibitem{2pucks-confirm}These results can be confirmed by a more
complete treatment of the problem with Newton's Laws and Kinematics.
\bibitem{negative-signs}I have been careful with my negative signs
here, but this largely unnecessary so long as students can identify
the direction of energy flow, which is explicitly indicated in their
diagrams and graphs. The negative signs merely reinforce what they
should already understand conceptually.
\bibitem{extension-to-modeling}His approach is an extension of the
energy units in the modeling method of physics instruction, whose
materials include more details such as lesson plans, worksheets and
tests. See reference \cite{modeling}.
\end{thebibliography}
\end{document}